# Spin Correlations in the Geometrically Frustrated $R$BaCo$_4$O$_7$ Antiferromagnets: Mean-Field Approach and Monte-Carlo Simulations

D.D. Khalyavin,[1] P. Manuel,[1] J.F. Mitchell,[2] and L.C. Chapon[1]

[1]*ISIS facility, Rutherford Appleton Laboratory-STFC, Chilton, Didcot, Oxfordshire, OX11 0QX, UK.*
[2]*Materials Science Division, Argonne National Laboratory, Argonne, Illinois 60439, USA.*
(Dated: August 4, 2010)

Spin correlations in the geometrically frustrated $R$BaCo$_4$O$_7$ compounds, usually described as an alternating stacking of Kagome and triangular layers on a hexagonal lattice, have been studied by mean-field approach and by Monte-Carlo simulations. The behavior of the system was modelled with an isotropic Heisenberg Hamiltonian as a function of the relevant parameter $J_{out}/J_{in}$, representing the ratio between exchange integrals inside the Kagome layers, $J_{in}$, and between Kagome and triangular layers, $J_{out}$. This ratio can be varied in real systems by appropriate chemical substitutions. At the mean-field level, long-range magnetic order with the wave vector at the $K$-point of symmetry ($\mathbf{k}=\mathbf{a}^*/3+\mathbf{b}^*/3$) has been found for $J_{out}/J_{in} > 0.7$. Below this value, the dominant Fourier modes are completely degenerate in the entire Brillouin zone. The Monte-Carlo simulations revealed that the long-range ordered configuration found in the mean-field calculations becomes the ground state of the system for $J_{out}/J_{in} > 1.5$. Below this critical ratio, quasi one-dimensional magnetic ordering along the $c$-axis, involving spins of the triangular sublattice was observed. The correlations in the $(ab)$ plane were found to have a short-range $120^o$ character with correlation length dependent on $J_{out}/J_{in}$.

## I. INTRODUCTION

Understanding complex magnetic behaviour in geometrically frustrated antiferromagnets has been the subject of numerous investigations over the last decades. The presence of frustrated spins often leads to the suppression of long-range magnetic ordering and promotes short-range correlations due to fluctuations between nearly or totally degenerated ground states.[1] These fluctuations, in particular for "spin-liquid" regimes, extend up to very low temperatures (far below the Curie-Weiss temperature) resulting in unusual static and dynamic magnetic states. Many complex rare-earth and transition metal oxides crystallize with structures whose motif has the potential to exhibit geometrical frustration.[2] Well known examples are transition-metal spinels,[3] jarosites[4] and rare-earth titanates and stannates with pyrochlore structure.[5] In these compounds, magnetic cations form two- or three-dimensional sublattices which can be decomposed in triangular or tetrahedral units for which three-fold symmetry is fundamentally incompatible with antiferromagnetic (AFM) ordering, promoting large ground-state degeneracy.

Recently, a new potentially frustrated structure has been reported for $R$BaCo$_4$O$_7$ cobalt oxides.[6,7] The hexagonal lattice of these compounds ($P6_3mc$ symmetry) consists of corner sharing CoO$_4$ tetrahedra arranged in alternating Kagome and triangular layers (Fig. 1a) stacked along the $c$-axis. The sublattice formed by cobalt ions corresponds to a new exchange topology (Fig. 1b) where geometrical frustration is inherent to the presence of both trigonal bipyramids and triangular clusters.[8] The former is an example of a new building unit which has never been observed before in other frustrated systems.

Considering an isotropic Heisenberg Hamiltonian,

$$H = -\frac{1}{2}\sum_{i,j} J_{i,j} \mathbf{S}_i \mathbf{S}_j \tag{1}$$

where the spins $\mathbf{S}_i$ and $\mathbf{S}_j$ interacting with energy $J_{ij}$ are represented by classical three-component unit vectors (the spin magnitude is included in the exchange parameters $J_{ij}$ which are positive for ferromagnetic interactions and negative for antiferromagnetic ones), the level of geometric frustration[9] in a cluster can be defined with a constraint function, $F_c = E_f / E_b$, where $E_f$ is the energy of the ground state of the frustrated spin arrangement and $E_b$ is the "basis" energy of the hypothetical spin arrangement in which all interactions between pairs of spins are satisfied. $F_c$ ranges from -1 (non frustrated case) to +1 (fully frustrated case). For canonical frustrated units like triangle and tetrahedron, non-collinear configurations satisfying $\sum_i \mathbf{S}_i = 0$ condition are expected to be the ground states. These are for instance well known $120^o$ (for triangle) and $109^o$ (for tetrahedron) configurations with $F_c$ equals to -1/2 and -1/3, respectively. In the latter case, the $109^o$ configuration is not a unique ground state, since any spin arrangement with $\sum_i \mathbf{S}_i = 0$ has equally low energy. In contrast, for a trigonal bipyramid with a unique exchange parameter between all nearest neighbours, the total zero moment is not the only condition for the ground state. As shown previously,[8] only configurations with parallel apical spins can minimize the exchange energy in the cluster (Fig. 1c). This additional condition restricts the number of ground states but does not reduce their number down to a single specific configuration as in the case of triangle. The internal degrees of freedom which can be varied without changing the energy of a bipyramidal cluster can be presented as two angular parameters indicated in Fig. 1c by $\theta$ and $\varphi$. Examples of some ground-state configurations with three-dimensional and planar spin arrangements are given in Figure 1d. The energy of these configurations is $E_f = 3.5J$ which yields to the level of geometric frustration $F_c = $ -7/18. Thus, the bipyramid can be classified between the triangle and the tetrahedron in the "hierarchy of frustrated units".

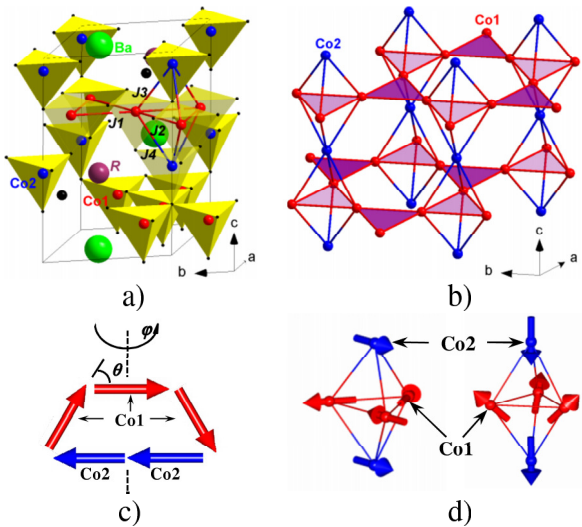

FIG. 1: (Color online) (a) Schematic representation of the hexagonal $P6_3mc$ crystal structure of $R$BaCo$_4$O$_7$ as a corner-shared CoO$_4$ tetrahedral network. Cobalt ions in Kagome and triangular layers are denoted as Co1 (red) and Co2 (blue), respectively. Four nonequivalent superexchange paths. $J_1$, $J_2$, $J_3$ and $J_4$ are shown as solid lines. (b) Extended view of the exchange topology demonstrating corner-sharing triangular (dark shadow) and bipyramidal (with base triangles as light shadow) units. Note that dark and light triangles together form the Kagome layers. (c) Schematic representation of ground state spin configuration for a bipyramidal unit with equivalent in-plane and out-of-plane interactions. Internal degrees of freedom which can be varied without changing the energy of the cluster are indicated by angular parameters $\theta$ and $\varphi$ (the first defines angles between spins in the planar geometry, when all spins in both Co1 and Co2 positions are parallel to a same plane; the second is responsible for taking the Co1 spins out of the planar configuration) (d) Example of two ground state configurations with spatial (right) and planar (left) spin arrangements.

In real materials with triangular planar lattices, the non-collinear $120^o$ structure occurs quite commonly[4,10-12] while $109^o$ long-range ordered spin configuration has been found only for one pyrochlore-like topology in FeF$_3$.[13] Materials with tetrahedrally-based lattices usually adopt short-range correlated phases like spin ice or spin glass.[2] The situation with $R$BaCo$_4$O$_7$ is not clear yet. Several neutron diffraction studies have been reported with contradictory conclusions. Soda et al.[14] observed only short-range correlations in YBaCo$_4$O$_7$ single crystal and discussed the result based on model where the Kagome and triangular layers are decoupled and form spin configurations close to $120^o$. In contrary, a long range magnetic ordering with intricate magnetic structure has been reported for polycrystalline sample of this composition below $T_N$ 110 K.[15] Above $T_N$, diffuse scattering was observed in a wide temperature range for both polycrystalline and single crystal samples and successfully modelled using a Monte-Carlo method.[8] One central issue is that $R$BaCo$_4$O$_7$ cobaltites experimentally studied so far have shown a lower symmetry than that of the ideal hexagonal structure (trigonal $P31c$ or orthorhombic $Pna2_1$), which raise questions about possible modifications of certain exchange integrals and the associated lifting (even partial) of the frustration. Moreover, a high resolution diffraction study[18] revealed that the oxygen content greatly influences the structural and magnetic behaviour in those cobaltites, a parameter that complicates further the analysis and comparison of various experimental data.

In the present work, we consider spin correlations in the $R$BaCo$_4$O$_7$ cobaltites based on a mean-field approach and Monte Carlo simulations. We discuss the cases for the hexagonal $P6_3mc$ and trigonal $P31c$ symmetry but leave out the orthorhombic $Pna2_1$ one since the latter does not preserve the three-fold symmetry of the bipyramidal and triangular units and therefore effectively removes the frustration. First, the exchange parameters are analyzed using extended Huckel tight binding (EHTB) calculations which show that the nearest neighbour interactions are very strong for these compositions. Relative strengths of the different exchange integrals are estimated for both hexagonal and trigonal phases based on the structural parameters derived experimentally. Then, mean-field calculations are performed for different sets of the exchange interactions. We show that for the hexagonal structure with nearly equivalent exchange parameters, long-range magnetic ordering can be stabilized with a tripled unit cell in the $(ab)$ plane. This situation is different from the "classical" frustrated Kagome and pyrochlore lattices where no long-range order is predicted in the mean-field approximation due to the presence of dispersionless modes.[17] The Monte-Carlo simulations are conducted using two exchange parameters in the system representing respectively interactions inside the Kagome layers and between Kagome and triangular layers. Varying the ratio between these parameters, one observes a gradual transition from a quasi one-dimensional ordering of spins in the triangular lattice to the three-dimensional ordered configuration obtained in the mean-field calculations. Finally, based on the Monte-Carlo simulations, neutron scattering cross sections for different cuts of the reciprocal space were calculated and plotted as two dimensional 2D maps. These plots provide information about diffuse magnetic scattering necessary to interpret neutron diffraction experiments.

## II. EXCHANGE INTERACTIONS

As mentioned above, the structural motif of the $RBaCo_4O_7$ cobaltites consists of corner sharing $CoO_4$ tetrahedra arranged in alternating Kagome and triangular layers (Fig. 1a). The strongly coupled units are expected to be the neighbor tetrahedra linked by a common oxygen ion. The superexchange interactions responsible for the spin coupling are known to be determined by both geometrical factors related to a crystal lattice and the electronic structure of the solid.[18] In the case of the parent hexagonal $P6_3mc$ structure, two possibilities can be distinguished: when atoms occupy the highly symmetric positions and when some atomic displacements allowed by the symmetry are present. For the former situation, all the superexchange paths are equivalent and there is a single exchange parameter, $J$, whereas for the latter one, four parameters, $J_1$, $J_2$, $J_3$ and $J_4$, should be considered (Fig. 1a); the first two inside Kagome planes (labelled as in-plane) and the last two between Kagome and adjacent triangular layers (labelled as out-of-plane). It is not straightforward to estimate the sign and strength of the exchange interactions in these compositions using the simple Goodenough-Kanamory-Anderson rules,[19] therefore a more semi-quantitative approach was adopted by spin-dimer analysis based on molecular electronic structure calculation.[18,20] The analysis was performed with the aid of the package *CAESAR 2.0*.[21]

Following Whangbo, Koo and Dai[18] it is convenient to write exchange parameter as $J = J_F + J_{AF}$, where the ferromagnetic term $J_F > 0$ and antiferromagnetic $J_{AF} < 0$. In general, $J_F$ is small so that the spin exchange cannot be ferromagnetic unless the $J_{AF}$ either vanishes or is very small in magnitude by symmetry. The latter can be approximated as $J_{AF} = (\Delta e)^2 / U_{eff}$, where $\Delta e$ is the energy split due to orbital-orbital interaction and $U_{eff}$ is an effective one site repulsion which is nearly constant for a given magnetic system. Thus, trends in $J_{AF}$ are well approximated by those in the corresponding $(\Delta e)^2$. The energy split can be obtained from electronic structure calculations based on EHTB method, performed for an appropriate structural fragment representing a spin-dimer.[18,20] These calculations were done for both hexagonal $P6_3mc$ and trigonal $P31c$ structures, taking as electronic basis $4s$, $4p$ single-zeta and $3d$ double-zeta Slater type orbitals of cobalt and $2s$, $2p$ double-zeta STO of oxygen. The double-zeta STO, $\chi_\mu$, is defined as linear combination of two exponential functions

$$\chi_\mu \propto r^{n-1} \left[ c_1 e^{-\xi_1 r} + c_2 e^{-\xi_2 r} \right] Y(\Theta, \Phi)$$

where $n$ is the principal quantum number, $\xi_i$ are the exponents and $Y(\Theta, \Phi)$ is the appropriate spherical harmonics. The exponents $\xi_1$ and $\xi_2$ in the radial part of STO describe the compression and diffuseness of the orbital, respectively.[22] In the case of a single-zeta STO, $c_2 = 0$. The valence state ionization potentials of these orbitals used for the construction of the matrix representation of the effective one-electron Hamiltonian.

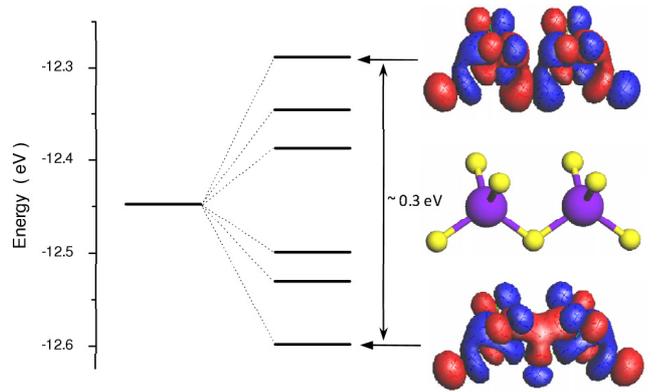

FIG. 2: (Color online) Splitting of $t_{2g}$ energy level due to orbital-orbital interactions in $CoO_4$-$CoO_4$ dimer. Right side demonstrates antibonding (top) and bonding (bottom) molecular orbitals corresponding to the largest energy splitting. In between the molecular orbitals, the structural fragment representing the spin dimer is shown.

$H_{\mu\nu} = \langle \chi_\mu H_{eff} \chi_\nu \rangle$) were taken to be -9.21, -5.29, -13.18, -32.3 and -14.8 eV, respectively.[23] At first, the calculations were done for the ideal hexagonal structure where all tetrahedral units are perfectly symmetric (equivalent superexchange paths). For simplicity, we considered only contribution to $\Delta e$ coming from $t_{2g}$ orbitals which are expected to be magnetic in the case of $Co^{2+}$ ($e_g^4 t_{2g}^3$) ions. We use $t_{2g}$ and $e_g$ notations for the molecular orbitals of the dimer with the largest weighting coefficients of the corresponding $3d$ atomic orbitals of cobalt. The contribution caused by $e_g$ orbitals, one of which is magnetic for $Co^{3+}$ ($e_g^3 t_{2g}^3$) ions, is negligible because of their small interaction and low concentration of $Co^{3+}$. Also, possible charge ordering between $Co^{2+}$ and $Co^{3+}$ proposed in the early reports[6,7] was subsequently disproved by the detailed structural investigations.[15,18] A part of the electronic diagram demonstrating the energy splitting due to interaction of $t_{2g}$ molecular orbitals in the $CoO_4$-$CoO_4$ dimer is shown in Figure 2. The large value of $(\Delta e)^2 \sim 140000$ meV$^2$ caused by strong orbital overlapping points to a dominant role of the $J_{AF}$ term and predicts very strong antiferromagnetic interactions in agreement with experimental observations.[15,24]

As mentioned above, actually, the hexagonal $P6_3mc$ symmetry allows some atomic shifts to be present which, in the most general case, split the single exchange parameter of the ideal structure into a set of four nonequivalent ones. The absence of reliable structural data related to the hexagonal phase, which might only be stabilized at high temperature but not yet observed, makes it difficult to estimate the dispersion of the exchange parameters introduced by these shifts. The simplest way to quantify this effect is to use the mode decomposition analysis described for $TmBaCo_4O_7$ in Ref.[25] Starting from the structural parameters refined in the trigonal $P31c$ space group, the distortions were decomposed in terms of sym-metry adapted displacement modes transforming like basis functions of the $\Gamma_1$ and $\Gamma_3$ irreducible representations (irreps) of the $P6_3mc$ space group. The distortions related to the $\Gamma_1$ irrep are secondary in the $P31c$ structure and their presence is expected in the $P6_3mc$ one as well. Thus, by putting the amplitudes of the $\Gamma_3$ modes to zero and generating new coordinates involving only $\Gamma_1$ displacements, we can estimate different values of $J$s in the "distorted" hexagonal structure. The relevant analysis using *ISODISPLACE* software[26] shows that all the exchange parameters are very close to those found for the ideal case. The largest difference is less than 5%. Thus, with a good approximation we can use just a single exchange parameter for the hexagonal structure.

The spin-dimer analysis was then performed for the $P31c$ trigonal phase using the structural data from Ref.[25] The trigonal distortions do not introduce additional splitting of the exchange parameters and like in the case of the hexagonal symmetry there are four $J$s not constrained by symmetry to be equal (Fig. 1a). The analysis revealed that the difference between the two in-plane interactions, $J_1$ and $J_2$, is again very small $\sim 4\%$ ($J_2 = 0.96J_1$). To simplify the evaluation of the out-of-plane exchange parameters coupling crystallographically non-equivalent cobalt ions, we did not take into account the fact that $t_{2g}$ orbitals are not degenerated in the presence of the trigonal distortions. With this simplification, the estimated values for $J_3$ and $J_4$ were found to be $0.74J_1$ and $0.82J_1$, respectively. Based on this result, one can conclude that the symmetry lowering down to trigonal $P31c$ affects mainly the out-of-plane interactions. Experimentally, a change of the trigonal distortions by external pressure can essentially modify these interactions. Selective substitutions such as for instance $Al^{3+}$ or $Zn^{2+}$ which preferably occupy triangular or Kagome sublattices,[27–30] respectively are also an effective way to alter the relative strengths of interactions inside Kagome and between Kagome and triangular layers. In the following consideration, we therefore mainly restrict our discussion to the idealized case of two exchange parameters; one in-plane, $J_{in} = J_1 = J_2$ and one out-of-plane, $J_{out} = J_3 = J_4$.

## III. MEAN-FIELD APPROACH

Next, we investigate the ordered magnetic phases in the $RBaCo_4O_7$ system, within mean-field approximation and classical spin Hamiltonian (1), which in the case of an extended solid with a crystalline structure can be written as:

$$H = -\sum_{i,j} \sum_{\boldsymbol{R},\boldsymbol{R'}} J_{i,j}^{\boldsymbol{R},\boldsymbol{R'}} \boldsymbol{S}_i^{\boldsymbol{R}} \boldsymbol{S}_j^{\boldsymbol{R'}} \qquad (2)$$

where $J_{i,j}^{\boldsymbol{RR'}}$ is the exchange integral between spin $\boldsymbol{S}_i$ in the primitive cell with a lattice vector $\boldsymbol{R}$ and spin $\boldsymbol{S}_j$ in the cell with a lattice vector $\boldsymbol{R'}$. We follow the approach used by Reimers, Berlinsky and Shi.[17] The translation symmetry implies that the magnetic order parameters, $\sigma_i^{\boldsymbol{R}}$, representing the mean value of the spins on sites $i$ in a cell with lattice vector $\boldsymbol{R}$, can be expressed as a Fourier series:

$$\sigma_i^{\boldsymbol{R}} = \sum_{\boldsymbol{k}} \sigma_i^{\boldsymbol{k}} e^{-i\boldsymbol{k}\boldsymbol{R}} \qquad (3)$$

where the sum is taken over all propagation vectors, $\boldsymbol{k}$, involved in the transition to the ordered state. Analogously, the spin-spin interaction matrix, $\xi_{ij}$, in terms of its Fourier components can be written as:

$$\xi_{ij} = -\sum_{\boldsymbol{R'}} J_{ij}^{\boldsymbol{R'}} e^{-i\boldsymbol{k}\boldsymbol{R'}} \qquad (4)$$

Diagonalization of quadratic part of the mean-field free-energy requires transforming into the normal modes of the system and solving the eigenvalue problem:

$$-\sum_j \left[ \sum_{\boldsymbol{R'}} J_{ij}^{\boldsymbol{R'}} e^{-i\boldsymbol{k}\boldsymbol{R'}} \right] \sigma_j = \lambda(\boldsymbol{k})\sigma_i \qquad (5)$$

The vector $\boldsymbol{k}$ minimizing the lowest eigenvalue, $\lambda_{min}$, of the interaction matrix (4), for a given set of exchange parameters in the system is the propagation vector of the magnetic structure occurring just below the transition temperature ($T_{cr} = -\lambda_{min}/3k_B$) from the disordered state. The corresponding eigenvector, $\boldsymbol{\sigma}$, yields the Fourier coefficients, $\sigma_i^{\boldsymbol{k}}$, of the spin configuration with the exchange energy $\lambda_{min}$ per unit cell. This spin configuration is usually referred to as the first ordered state. If the eigenvector keeps the magnitude of the spins for each site constant ($|\sigma_i^{\boldsymbol{k}}| = \frac{1}{\sqrt{N}}, i = 1, 2 \ldots N$; $N$ is number of spins in a unit cell) then, following Luttinger and Tisza theorem,[31] the first ordered state is simultaneously the ground state of the system. If this condition is not satisfied then the symmetry and periodicity of the ground state can be different and additional transitions can take place below the critical temperature of the magnetic ordering. In the case when equivalent minima occur for more than a single wave vector, then the first ordered state is expected to be a linear combination of the corresponding eigenvectors.

The mean-field calculations for Kagome and pyrochlore systems have been done by Reimers, Berlinsky and Shi.[17] An unusually high degree of degeneracy has been found when taking only antiferromagnetic nearest-neighbour interactions into account. In this case, some of the Fourier modes are completely degenerate for all wave vectors in the first Brillouin zone and no long-range order is possible for both systems. Let us consider now the ordered states in $RBaCo_4O_7$. We will include only interactions between nearest neighbours, which according to the spin-dimer analysis are extremely strong in these compositions. Using the structural parameters summarised in Table I, the exchange matrix (4) for a general case of the four non-equivalent exchange parameters can be written as:

$$-\begin{pmatrix}
0 & 0 & J_4e^{ik_y} & J_4 & J_4e^{i(k_x+k_y)} & J_3e^{ik_x} & J_3 & J_3e^{i(k_x+k_y)} \\
0 & 0 & J_3e^{ik_y} & J_3 & J_3e^{i(k_x+k_y)} & J_4e^{i(k_x+k_z)} & J_4e^{-ik_z} & J_4e^{i(k_x+k_y-k_z)} \\
J_4e^{-ik_y} & J_3e^{-ik_y} & 0 & J_1+J_2e^{-ik_y} & J_1+J_2e^{ik_x} & 0 & 0 & 0 \\
J_4 & J_3 & J_1+J_2e^{ik_y} & 0 & J_1+J_2e^{i(k_x+k_y)} & 0 & 0 & 0 \\
J_4e^{-i(k_x+k_y)} & J_3e^{-i(k_x+k_y)} & J_1+J_2e^{-ik_x} & J_1+J_2e^{-i(k_x+k_y)} & 0 & 0 & 0 & 0 \\
J_3e^{-ik_x} & J_4e^{-i(k_x-k_z)} & 0 & 0 & 0 & 0 & J_1+J_2e^{-ik_x} & J_1+J_2e^{ik_y} \\
J_3 & J_4e^{ik_z} & 0 & 0 & 0 & J_1+J_2e^{ik_x} & 0 & J_1+J_2e^{i(k_x+k_y)} \\
J_3e^{-i(k_x+k_y)} & J_4e^{-i(k_x+k_y-k_z)} & 0 & 0 & 0 & J_1+J_2e^{-ik_y} & J_1+J_2e^{-i(k_x+k_y)} & 0
\end{pmatrix} \quad (6)$$

Diagonalization of the matrix for different values of $J$s and $\boldsymbol{k}$ was done numerically with the aid of the program *ENERMAG*.[32] For each set of exchange integrals, we have investigated the dispersion relations by varying the $\boldsymbol{k}$ vector on the surface and inside the first Brillouin Zone. Although the hexagonal $P6_3mc$ and trigonal $P31c$ symmetries allow four exchange parameters to be nonequivalent, we will first analyse two simplified situations. i) when all the exchange integrals are identical. $J_1 = J_2 = J_3 = J_4 = J$ and ii) when in-plane interaction $J_{in} = J_1 = J_2$ is not equivalent to the out-of-plane one $J_{out} = J_3 = J_4$. The former case corresponds to the idealized parent hexagonal structure, the latter to the idealized trigonal one. The obtained dispersion relations along some lines of symmetry are presented in Figures 3 and 4. The result revealed that in the most symmetric case with the single exchange parameter in the system, the first ordered state is not degenerate. The wave vector at the $K$-point of symmetry ($\boldsymbol{k}=\boldsymbol{a}^*/3+\boldsymbol{b}^*/3$) minimizes the lowest eigenvalue ($\lambda_{min} = -3 \mid J \mid$) of the exchange matrix (6). The related eigenvector is given in Table II (Ph1). The spin arrangement in the corresponding ordered state is shown in the insert of Figure 4 and Figure 5 presents the structure as two separate layers with the cobalt fractional coordinates $z < 0.5$ and $z > 0.5$. In this magnetic structure, the equatorial spins (Co1) in the base triangles of the bipyramidal units are parallel to each other and antiparallel to the apical spins (Co2). The magnitude of the latter is twice as large as that of the equatorial spins so the sum over all the magnetic moments in the cluster does not equal zero as it is expected for the ground state configuration. On the contrary, the spins in the triangular units linking different bipyramidal clusters are arranged to create $120^o$ configuration with zero total moment and uniform chirality within the ($ab$) plane. In the neighbouring planes, the chirality has opposite sign (Fig. 5).

Thus, the calculations demonstrate the crucial difference between the ideal bipyramidal topology of $R$BaCo$_4$O$_7$ and tetrahedral pyrochlore lattice where no long-range order is predicted in the mean-field approximation. As previously mentioned, all compositions of the $R$BaCo$_4$O$_7$ series investigated up to now have symmetry lower than hexagonal. Due to stereochemical ef-fects the actual symmetry is either orthorhombic $Pna2_1$ or trigonal $P31c$. The latter is of special interest since it preserves the three-fold axis of the bipyramidal and triangular units. As it was shown based on EHTB calculations, the symmetry lowering down to trigonal $P31c$ affects mainly the out-of-plane interactions. Dispersion relations for different ratios of $J_{out}/J_{in}$ are presented in Figure 3. Decreasing the magnitude of the out-of-plane interactions results in an increase of the energy of the lowest eigenvalue at the $K$ and $M$-points. However, the difference in energy at these two points is always very small ($\sim 1\%$). The important feature of the obtained dispersion curves is the presence of the two low lying dispersionless branches whose energy does not depend on the out-of-plane interaction and is entirely determined by the in-plane one ($\lambda_{deg} = -2\,|\,J_{in}\,|$). At the critical value of $J_{out}/J_{in} \sim 0.7$, these dispersionless branches become the lowest eigenvalues of the exchange matrix (6). The presence of these degenerate modes prevents long-range magnetic ordering in compositions where the ratio $J_{out}/J_{in}$ is below the critical one. The situation becomes very similar to that discussed by Reimers, Berlinsky and Shi[17] for the pyrochlore lattice. Thus, this result in combination with the spin dimer analysis, points to the fact that symmetry lowering down to trigonal $P31c$ can effectively contribute to a destabilisation of the long-range magnetic ordering in the $R$BaCo$_4$O$_7$ compositions.

TABLE I: Structural parameters for Co ions in the hexagonal $P6_3mc$ space group. The coordinates are given for the highly symmetric positions in the primitive unit cell with the lattice translation $(0,0,0)$.

| Atom(site) | No | $x,y,z$ | Neighbour No (lattice translation, $\boldsymbol{R}$) |
|---|---|---|---|
| Co1(2a) | 1 | 0,0,7/16 | 3(0,-1,0), 4(0,0,0), 5(-1,-1,0) 6(-1,0,0), 7(0,0,0), 8(-1,-1,0) |
| Co1(2a) | 2 | 0,0,15/16 | 3(0,-1,0), 4(0,0,0), 5(-1,-1,0) 6(-1,0,1), 7(0,0,1), 8(-1,-1,1) |
| Co2(6c) | 3 | 1/6,5/6,11/16 | 1(0,1,0), 2(0,1,0), 4(0,0,0) 4(0,1,0), 5(0,0,0), 5(-1,0,0) |
| Co2(6c) | 4 | 1/6,1/3,11/16 | 1(0,0,0), 2(0,0,0), 3(0,0,0) 3(0,-1,0), 5(0,0,0), 5(-1,-1,0) |
| Co2(6c) | 5 | 2/3,5/6,11/16 | 1(1,1,0), 2(1,1,0), 3(0,0,0) 3(1,0,0), 4(0,0,0), 4(1,1,0) |
| Co2(6c) | 6 | 5/6,1/6,3/16 | 1(1,0,0), 2(1,0,-1), 7(0,0,0) 7(1,0,0), 8(0,0,0), 8(0,-1,0) |
| Co2(6c) | 7 | 1/3,1/6,3/16 | 1(0,0,0), 2(0,0,-1), 6(0,0,0) 6(-1,0,0), 8(0,0,0), 8(-1,-1,0) |
| Co2(6c) | 8 | 5/6,2/3,3/16 | 1(1,1,0), 2(1,1,-1), 6(0,0,0) 6(0,1,0), 7(0,0,0), 7(1,1,0) |

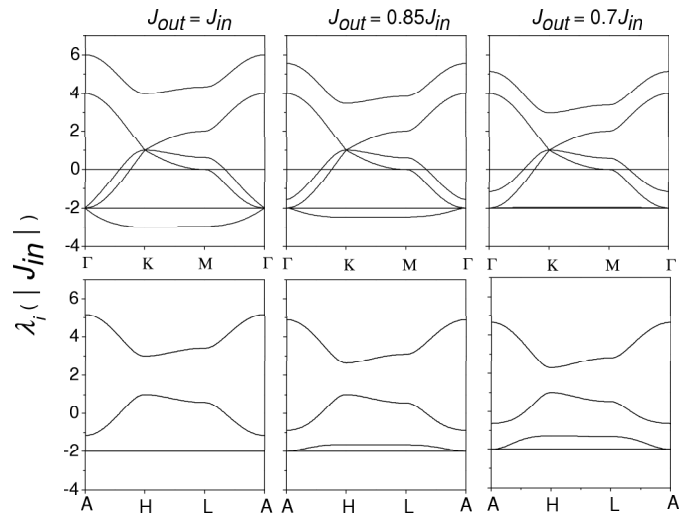

FIG. 3: Dispersion relations for the eigenvalues of the exchange matrix (6) for various $J_{out}/J_{in}$ ratios along some lines of symmetry of the first Brillouin Zone of the $P6_3mc$ space group.

TABLE II: Eigenvectors and periodicity of the ordered states obtained by minimization of the lowest eigenvalue of the exchange matrix (6) for different sets of the exchange integrals ($J$s).

| Phase/(representative set of $J$s) | $\mathbf{k}$ | Eigenvector$^a$ |
|---|---|---|
| Ph1/($J_1 = J_2 = J_3 = J_4 = -1$) | $\dfrac{\mathbf{a^*}}{3} + \dfrac{\mathbf{b^*}}{3}$ | $\dfrac{1}{\sqrt{7}}\left(\sqrt{2}e^{-\frac{2}{3}\pi i}, \sqrt{2}e^{-\frac{2}{3}\pi i}; \dfrac{1}{\sqrt{2}}e^{-\frac{\pi}{3}i}, \dfrac{1}{\sqrt{2}}e^{\frac{\pi}{3}i}, -\dfrac{1}{\sqrt{2}}; \dfrac{1}{\sqrt{2}}e^{-\frac{\pi}{3}i}, \dfrac{1}{\sqrt{2}}e^{\frac{\pi}{3}i}, -\dfrac{1}{\sqrt{2}}\right)$ |
| Ph2/($J_1 = J_2 = -1, J_3 = J_4 = 1$) | $\dfrac{\mathbf{a^*}}{3} + \dfrac{\mathbf{b^*}}{3}$ | $\dfrac{1}{\sqrt{7}}\left(\sqrt{2}e^{\frac{\pi}{3}i}, \sqrt{2}e^{\frac{\pi}{3}i}; \dfrac{1}{\sqrt{2}}e^{-\frac{\pi}{3}i}, \dfrac{1}{\sqrt{2}}e^{\frac{\pi}{3}i}, -\dfrac{1}{\sqrt{2}}; \dfrac{1}{\sqrt{2}}e^{-\frac{\pi}{3}i}, \dfrac{1}{\sqrt{2}}e^{\frac{\pi}{3}i}, -\dfrac{1}{\sqrt{2}}\right)$ |
| Ph12/($J_1 = J_2 = J_3 = -1, J_4 = 1$) | $\dfrac{\mathbf{a^*}}{3} + \dfrac{\mathbf{b^*}}{3}$ | $\dfrac{1}{\sqrt{7}}\left(\sqrt{2}e^{\frac{\pi}{3}i}, \sqrt{2}e^{-\frac{2}{3}\pi i}, \dfrac{1}{\sqrt{2}}e^{-\frac{\pi}{3}i}; \dfrac{1}{\sqrt{2}}e^{\frac{\pi}{3}i}, -\dfrac{1}{\sqrt{2}}; \dfrac{1}{\sqrt{2}}e^{\frac{2}{3}\pi i}, \dfrac{1}{\sqrt{2}}e^{-\frac{2}{3}\pi i}, \dfrac{1}{\sqrt{2}}\right)$ |
| POS/($J_1 = 2, J_2 = -2, J_3 = J_4 = -1$) | $\dfrac{\mathbf{a^*}}{3} + \dfrac{\mathbf{b^*}}{3}$ | $\dfrac{1}{\sqrt{3}}\,(0,0,1,1,1,0,0,0) + \dfrac{1}{\sqrt{3}}\,(0,0,0,0,0,1,1,1)^b$ |
| F/($J_1 = J_2 = J_3 = J_4 = 1$) | 0 | $\dfrac{1}{2\sqrt{2}}\,(1,1,1,1,1,1,1,1)$ |
| Fi/($J_1 = J_2 = 1, J_3 = J_4 = -1$) | 0 | $\dfrac{1}{2\sqrt{2}}\,(1,1,-1,-1,-1,-1,-1,-1)$ |
| AF/($J_1 = J_2 = J_3 = 1, J_4 = -1$) | 0 | $\dfrac{1}{2\sqrt{2}}\,(1,-1,1,1,1,-1,-1,-1)$ |

$^a$eigenvectors for $-k$ are complex conjugated to those for $k$

$^b$for the POS state, the spin configuration is expected to be a linear combination of the two eigenvectors with equivalent eigenvalues.

By varying $J_{in}$ and $J_{out}$ in the interval [-5, 5], the complete phase diagram for the simplified case of the two exchange parameters was obtained (Fig. 6a). The Fourier coefficients of the ordered states which can be sta-bilized in these conditions are presented in Table II. In the case of negative exchange interactions in the Kagome layers ($J_{in} < 0$), the magnetic state is either disordered due to the dominance of the dispersionless modes or ordered with the propagation vector ($\mathbf{k}=\mathbf{a^*}/3+\mathbf{b^*}/3$). The phase boundaries between these states are straight lines whose slope is determined by the ratio $J_{out}/J_{in}$ and are symmetric with respect to the sign of the out-of-plane interactions. It should be pointed out, that there is a difference between the spin configurations of the ordered phases with the $\boldsymbol{a^*}/3 + \boldsymbol{b^*}/3$ periodicity for negative (Ph1) and positive (Ph2) out-of-plane interactions (Table II). The difference relates to the mutual orientation of the apical spins in the bipyramidal clusters with respect to the ferromagnetic moments of the base triangles. The orientation is antiparalell for $J_{out} < 0$ (Ph1) and parallel for $J_{out} > 0$ (Ph2). Positive in-plane interactions ($J_{in} > 0$) stabilise non-frustrated ferrimagnetic (where spins in Kagome and triangular layers are antiparalell) and ferromagnetic configurations for $J_{out} < 0$ and $J_{out} > 0$, respectively.

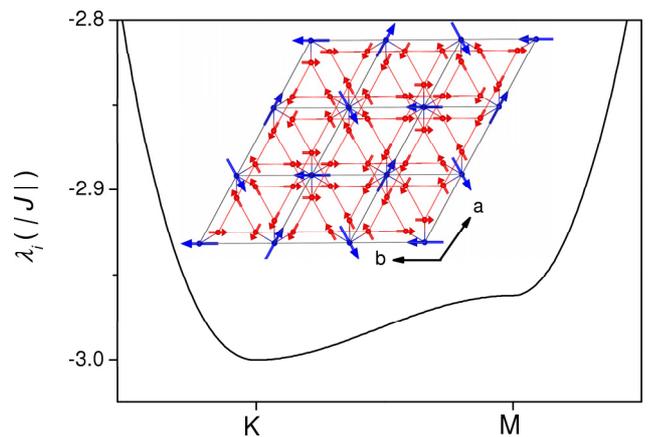

FIG. 4: (Color online) Dispersion curve demonstrating minimum at the $K$-point of symmetry for the case of equivalent in-plane and out-of-plane interactions. Inset shows the corresponding spin configuration (view along the $c$-axis). Spins in triangular and Kagome sublattices are shown by large (blue) and small (red) arrows, respectively.

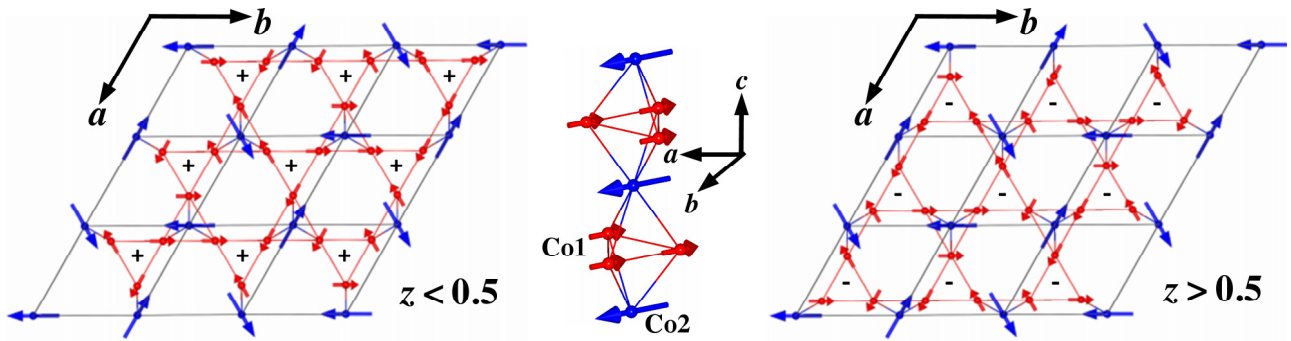

FIG. 5: (Color online) Spin configuration corresponding to the ordered state at $K$-point of symmetry for the case of equivalent in-plane and out-of-plane interactions. Spins in triangular and Kagome sublattices are shown by large (blue) and small (red) arrows, respectively. Two layers demonstrating ordering of the magnetic moments in the $(ab)$ plane for Co ions with fractional coordinates $z < 0.5$ (left) and $z > 0.5$ (right) are presented. In between, a structural fragment consisting of two bipyramidal clusters shared along the $c$ axis is shown. The "+" and "−" signs for the triangular units (shown in Figure 1b as dark shadow clusters) denote the sign of the chirality, defined as the rotation of spins as one goes around a triangle. The sign is positive/negative if anticlockwise/clockwise rotation takes place.

Finally, let us briefly discuss more complex situations, when there are more than two exchange parameters in the system. The following relations between the exchange parameters can be distinguished.

i) $J_1 = J_2 = J_{in} < 0$ and $J_3 \neq J_4$. If the out-of-plane interactions are not equivalent and in-plane interactions equal each other and are negative, the phase boundaries between the ordered and disordered states are again straight lines. The corresponding phase diagram is shown in Figure 6b where the out-of-plane interactions $J_3$ and $J_4$ are expressed in the units of the in-plane exchange parameter $J_{in}$. The diagram involves also a new ordered configuration denoted as Ph12 with a tripled unit cell in the $(ab)$ plane. This phase is stabilized when $J_3$ and $J_4$ have different signs. The corresponding magnetic structure can be presented as a simple combination of the phase Ph1 and Ph2 (Table II). The behaviour of the system can also be clearly presented if one introduces an average out-of-plane exchange parameter $J_{out\_aver}$. When $J_3$ and $J_4$ have the same signs, the $J_{out\_aver}$ parameter can be defined as: $J_{out\_aver} = (J_3 + J_4)/2$ and the phase diagram in terms of $J_{out\_aver}$ and $J_{in}$ coincides with that shown in Figure 6a ($J_{out\_aver}$ instead of $J_{out}$). When $J_3$ and $J_4$ have opposite signs it is convenient to define:

$J_{out\_aver} = (J_3 - J_4)/2$ and in this case, the phase diagram includes the ordered configurations Ph12 and AF (see Table II) instead of Ph1/Ph2 and F/Fi, respectively.

ii) $J_1 = J_2 = J_{in} > 0$ and $J_3 \neq J_4$. When in-plane interactions are positive only non-frustrated collinear ferromagnetic, ferrimagnetic or antiferromagnetic configurations with $\boldsymbol{k}=0$ are stable for any signs and values of $J_3$ and $J_4$ (not shown).

iii) $J_1 \neq J_2$ and $J_3 = J_4 = J_{out}$. For the case of non-equivalent in-plane interactions and equivalent out-of-plane ones, the phase diagrams are presented in Figure 6c ($J_{out} < 0$) and 6d ($J_{out} > 0$). The $J_1$ and $J_2$ are expressed in the units of $J_{out}$. The critical lines in the diagrams do not depend on the sign of $J_{out}$. The difference between these cases relates to which ordered configurations with the $\boldsymbol{k}=\boldsymbol{a^*}/3 + \boldsymbol{b^*}/3$ and $\boldsymbol{k}=0$ periodicities are stabilized (Ph1 or Ph2 and F or Fi). The phase boundary between the ordered and disordered states is not a straight line and depends also on the $J_1/J_2$ ratio. When $J_1 > 0$ and $J_2 < 0$, a partially ordered state (POS) can be stabilized in a wide range of the phase diagram. Spin ordering in the POS occurs only in Kagome layers, whereas the triangular sublattice remains disordered (Table II). It should be pointed out that besides the example considered here, one of the possible candidate for the exotic POS is $Gd_2Ti_2O_7$ pyrochlore, where the neutron diffraction experiments[33] revealed that 3/4 of the $Gd^{3+}$ spins ordered within Kagome planes, whereas the spins in triangular layers remained disordered. This phase was then found in the mean-field calculations by Cepas and Shastry.[34]

iv) $J_1 \neq J_2$ and $J_3 \neq J_4$. This most general situation can be presented in terms of the in-plane interactions $J_1$ and $J_2$ expressed in units of the average out-of-plane parameter, $J_{out\_aver}$, defined above. When both the out-of-plane interactions $J_3$ and $J_4$ are either positive or negative, the phase diagrams are the same as shown in Figure 6c $(J_{out\_aver} < 0)$ and 6d $(J_{out\_aver} > 0)$. If $J_3$ and $J_4$ have different signs, the phase Ph1 in the diagram shown in Figure 6c should be replaced by Ph12 and the phase Fi by the AF one.

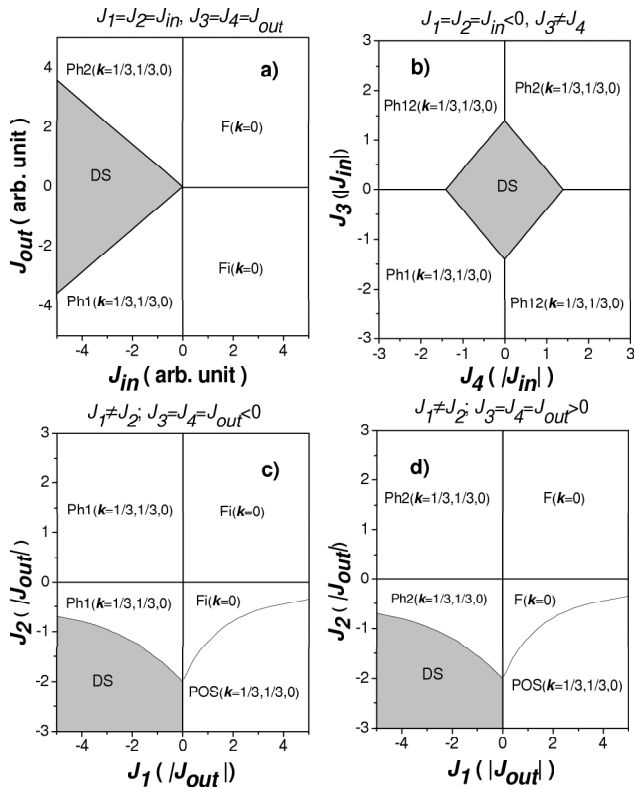

FIG. 6: Magnetic phase diagrams representing the stability of different states at some combinations of the exchange parameters. DS - disordered state, POS partially ordered state (ordering present only in Kagome layers), Ph1, Ph2 and Ph12 non-collinear configurations with tripled unit cell in the $(ab)$ plane (the coupling between adjacent Kagome and triangular layer along the $c$-axis is $----$ for Ph1, $++++$ for Ph2 and $++--$ for Ph12). Fi and F - ferrimagnetic and ferromagnetic phases, respectively.

## IV. MONTE-CARLO SIMULATIONS

To explore ground state configuration(s) of classical spins in the bipyramidal topology of $RBaCo_4O_7$, we used the direct Monte-Carlo algorithm[35,36] with Hamiltonian (2). The simulations have been done for a box containing 45 864 spins (a $21 \times 21 \times 13$ supercell) with open boundary conditions. The size of the box was chosen to estimate spin correlations up to tenth neighbours along the $c$-direction and up to ninth/eighteenth neighbours in the $(ab)$ plane for triangular/Kagome sublattice. The spins in the unit cells located on the surface of the box were not taken into account for calculating the correlations to minimise possible boundary effects. Only the simplified situation with two antiferromagnetic exchange parameters in the system, $J_{in} = J_1 = J_2$ and $J_{out} = J_3 = J_4$ was modelled. A simulated annealing was performed from a temperature $T/J = 100$ down to $T/J = 0.01$ ($J = J_{in}$ if $J_{in} > J_{out}$ and $J = J_{out}$ if $J_{in} < J_{out}$) with a cooling rate of 0.96 between the temperature steps. At each temperature, the program performed $10^4$ flips per spin using the standard Metropolis method.[35] For each particular ratio $J_{out}/J_{in}$, the simulation was run ten times to average residual fluctuations and to maximize the quality of scattering calculations. Based on the obtained spin configurations, the static spin-spin correlation functions ($\langle \mathbf{S}(0)\mathbf{S}(i_{\mathbf{R}}) \rangle = \sum_j \mathbf{S}_j \mathbf{S}_{j+i_{\mathbf{R}}}$) along different directions $\mathbf{R}$ ($i_{\mathbf{R}}$ is $i$-th neighbour along the direction $\mathbf{R}$) and the magnetic structure factor.

$$\mathbf{F}(\mathbf{q}) = \frac{\gamma r_0}{2N_x N_y N_z} \sum_i f_i(\mathbf{q}) \mathbf{S}_i e^{i\mathbf{q}\mathbf{R}_i}$$

where $\gamma$ is the neutron gyromagnetic ratio. $r_0$ is the free electron radius, $N_x$, $N_y$, $N_z$ are the supercell ($21 \times 21 \times 13$) dimension and $f_i(\mathbf{q})$ is the magnetic form factor for $Co^{2+}$ in the spherical approximation, for neutron scattering experiments have been derived. The neutron scattering intensity, $I(\mathbf{q})$, for a given scattering vector $\mathbf{q}$, was then calculated using $I(\mathbf{q}) = |\mathbf{e} \times \mathbf{F}(\mathbf{q}) \times \mathbf{e}|^2$, where $\mathbf{e}$ is the unit vector in the direction $\mathbf{q}$. Scattering maps for different cuts of the reciprocal space smoothed using a standard Gaussian smoothing are presented in (Fig. 7). Several control simulations with different box sizes and number of flips per spin proved negligible boundary effects and good thermal equilibration for the conditions described above.

The most symmetric case with a single exchange parameter in the system ($J_{in} = J_{out}$) has been discussed in the work by Manuel et al.[8] It was shown that the presence of corner-sharing bipyramids leads to quasi one-dimensional order along the $c$ axis in the apical Co2 site, whereas $120^o$ correlations of these spins in the $(ab)$ plane decay rapidly and do not exceed several unit cells as typically observed in spin liquids. The ground state condition $\sum_i \mathbf{S}_i = 0$, for each bipyramidal unit imposes the existence of analogous correlations in the Kagome layer as well (Fig. 1c). To demonstrate this let us introduce a resultant magnetic moment, $\mathbf{S}_\Sigma$, consisting of the three spins lying in the base of a trigonal bipyramid (Fig. 8a). One can see that the normalized correlation functions $\langle \mathbf{S}_\Sigma(0)\mathbf{S}_\Sigma(i_{\mathbf{R}}) \rangle$ along the $c$ direction and in the $(ab)$ plane are very similar to those calculated for the apical spins in the Co2 position (Fig. 8b-e). This effect is known for the pyrochlore lattice as well, where the ground state condition for tetrahedral units has been shown to result in a few long-range correlations along some high symmetry directions.[37] However, it should be pointed out that the one-dimensional ordering of spins in the Co2 site is a remarkable consequence of the unique exchange topology of $RBaCo_4O_7$ compounds and has no analogy in other known frustrated systems.

When the value of the out-of-plane interaction approaches zero, the system is expected to behave as a set of effectively decoupled Kagome layers. This scenario was discussed by Schweika. Valldor, and Lemmens[38] as the origin of the magnetic diffuse scattering signal observed for $Y_{0.5}Ca_{0.5}BaCo_4O_7$. The authors suggested that the cobalt ions in the triangular sublattice were in a low-spin diamagnetic state. However. this model was not supported by the single crystal neutron diffraction experiments performed for the $YBaCo_4O_7$ composition.[8]

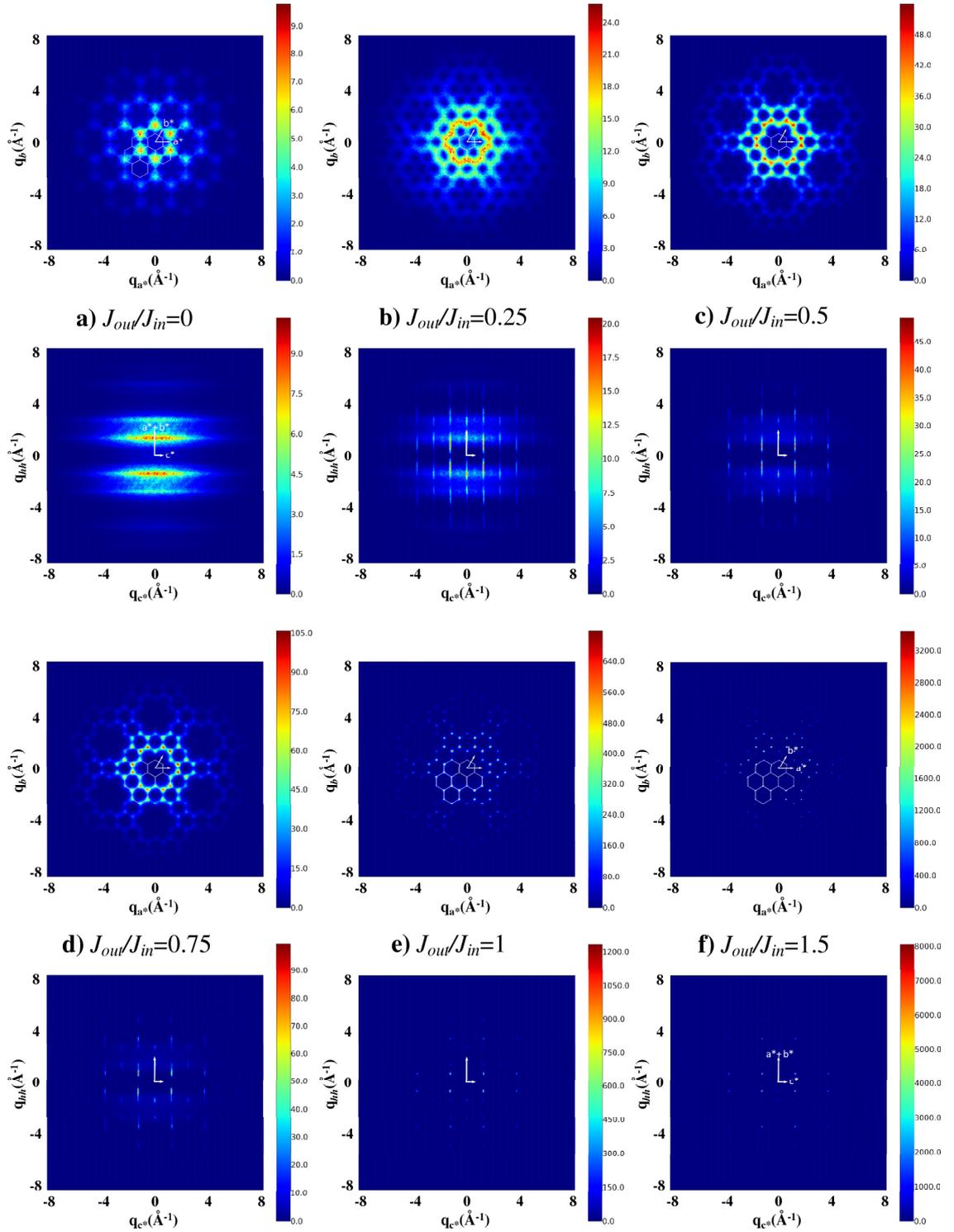

FIG. 7: (Color online) Simulated neutron scattering cross sections in the $(hk0)$ and $(hhl)$ reciprocal planes. Simulation done for the supercells obtained in the Monte-Carlo runs with different $J_{out}/J_{in}$ ratios.

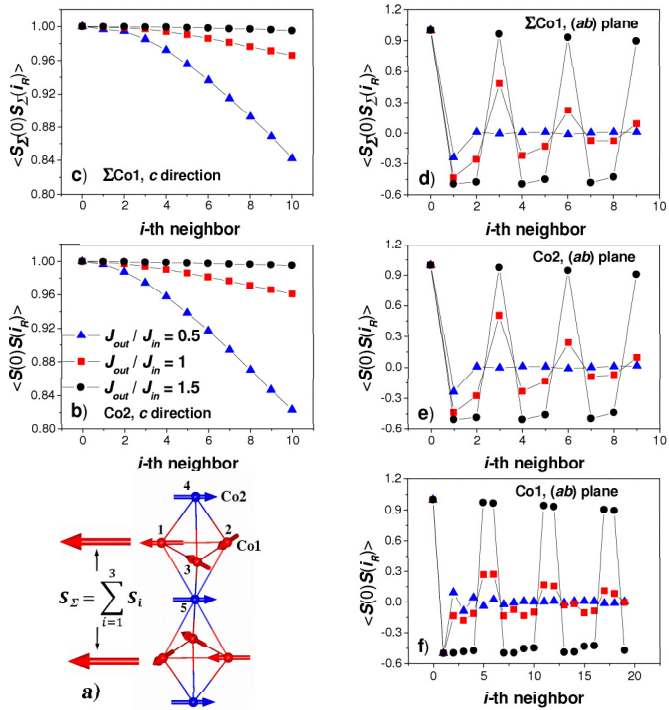

FIG. 8: (Color online) (a) A spin arrangement in a structural fragment demonstrating correlations of the resultant moments, $S_\Sigma$, in base triangles of bipyramidal units. (b-f) Spin-spin correlation functions for the Co1 and Co2 sites along the $c$ direction ($R = [001]$) and in the $(ab)$ plane ($R = [100]$, $[010]$ and $[110]$) evaluated for different $J_{out}/J_{in}$ ratios.

The observed magnetic diffuse scattering was successfully modelled only when spins in the triangular sublattice were participating actively in the exchange interactions. The simulated neutron diffraction patterns for the case of decoupled Kagome layers are shown in Figure 7a. The scattered intensity is not localised as sharp spots along the $c^*$ axis in contrast with the experimental data obtained for YBaCo$_4$O$_7$.[8]

For a finite value of $J_{out}$ even in the case of $J_{in} > J_{out}$, the correlation length along the $c$ axis is much larger than in the $(ab)$ plane (Fig. 8 and 9). This fact results in a strongly anisotropic simulated neutron scattering in the $(hhl)$ reciprocal plane (Fig. 7b-e); the intensity is well localized along the $c^*$ axis and much less along the $a^*+b^*$ direction. Apparently, one can always speak about "quasi"-long-range correlations along the $c$ direction, although modelling situations with very weak out-of-plane interactions is a difficult task because residual fluctuations become strong perturbations in this case. Increasing $J_{out}$ results in a gradual increase of the $120^\circ$ correlations in the $(ab)$ plane and, at the ratio $J_{out}/J_{in} = 1.5$, a transition to a three-dimensional long-range order takes place (Fig. 9). The indicator of the ordered state is a negligibly slow decay of the correlation function in all directions (Fig. 8) as well as sharp Bragg spots in the simulated diffraction patterns (Fig. 7f). The character of the correlations (Fig. 8b-f) and the diffraction picture (Fig 7f) both point to the tripled unit cell (wave vector $k = a^*/3 + b^*/3$) and are consistent with the spin configuration of the Ph1 obtained above in the mean-field calculations (Fig. 9(bottom), see also Fig. 5). In this configuration, all spins in a bipyramidal unit are parallel (Co1 site) or antiparallel (Co2 site) to some direction. In neighbouring bipyramids linked via a triangular unit in a tree-fold cluster, these directions make an angle of $120^o$.

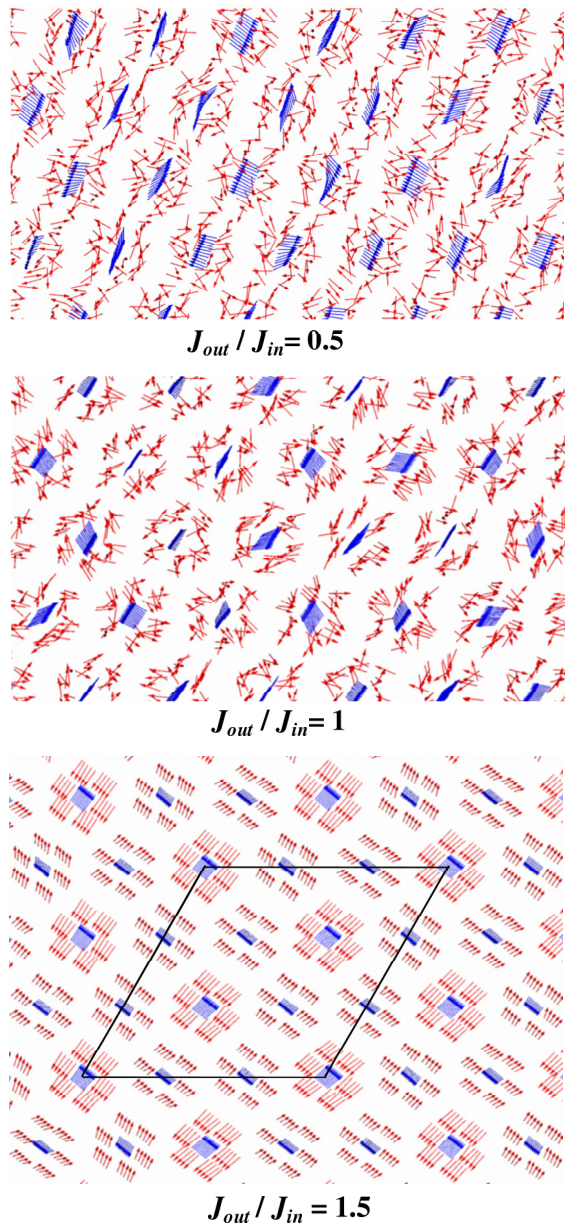

FIG. 9: (Color online) A part of $21 \times 21 \times 13$ supercell demonstrating spin arrangements obtained in Monte-Carlo simulations with different $J_{out}/J_{in}$ ratios. Perspective view along the $c$ axis. Ordered rows of spins running along the $c$ axis in the triangular sublattice are clearly visible in top and middle panels against a background of essentially disordered Kagome layers. For the case of $J_{out}/J_{in} = 1.5$ where the three dimensional long-range ordering takes place (bottom), a magnetic unit cell in the $(ab)$ plane is shown.

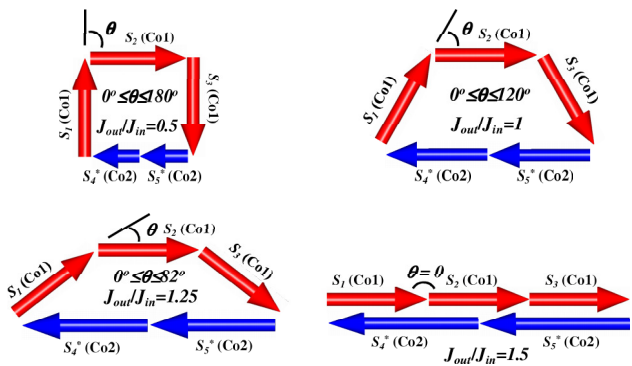

FIG. 10: (Color online) Ground state spin configurations of a bibyramidal cluster at different ratios $J_{out}/J_{in}$ (spins in the clusters are numerated ($\boldsymbol{S}_i, i = 1...5$) and for each of them a cobalt position (Co1/Co2) is given in brackets. To keep the $\sum_i \boldsymbol{S}_i = 0$ condition for the ground state, the values of spins in the triangular sublattice (Co2) were renormalized ($\boldsymbol{S}_i^* = \frac{J_{out}}{J_{in}}\boldsymbol{S}_i, i = 4, 5$), whereas the spins in the Kagome layers (Co1) have a fixed unit value.

The transition to the three-dimensional long-range order coincides with the disappearance of the internal degrees of freedom in the bipyramidal units as demonstrated in Figure 10, where few ground state configurations are shown for different values of the $J_{out}/J_{in}$ ratio. For clarity, only one degree of freedom indicated by the angular parameter $\theta$ is shown (planar configurations). To present these configurations we used the fact that renormalizing apical Co2 spins is equivalent to changing the $J_{out}$ exchange parameter. In this case, the Hamiltonian of the cluster can be written in a form similar to that given in Ref.[8]

$$ H = -\frac{J_{in}}{2}\left[(\boldsymbol{S}_1 + \boldsymbol{S}_2 + \boldsymbol{S}_3 + \boldsymbol{S}_4^* + \boldsymbol{S}_5^*)^2 - 2\boldsymbol{S}_4^*\boldsymbol{S}_5^* - C\right] $$

where $\boldsymbol{S}_i^*$ are renormalized apical spins ($\boldsymbol{S}_i^* = \frac{J_{out}}{J_{in}}\boldsymbol{S}_i$) and $C$ is a constant ($C = \boldsymbol{S}_1^2 + \boldsymbol{S}_2^2 + \boldsymbol{S}_3^2 + \boldsymbol{S}_4^{*2} + \boldsymbol{S}_5^{*2}$). This expression clearly shows that the minimum of the energy corresponds to zero total moment and parallel apical spins. Below the critical ratio $J_{out}/J_{in} = 1.5$, the angular parameter $\theta$ is not fixed in the ground state. The range of values over which this parameter can vary without changing energy of the cluster depends on the $J_{out}/J_{in}$ and decreases with increasing this ratio. For $J_{out}/J_{in} = 1.5$ the ground state conditions require all spins in the base triangle (Co1 position) to be antiparallel to the apical spins (Co2 position). Thus, the only possible value of $\theta$ is zero and therefore this configuration is not degenerated.

As mentioned above, spin correlations in the Kagome layers can be expressed using effective moments $\boldsymbol{S}_\Sigma$ which are correlated in the same way as spins in the triangular sublattice (Fig. 8b-e). Considering each spin in the

Kagome layers (Co1 position) separately, one can distinguish two components: parallel and normal to $\boldsymbol{S}_\Sigma$. The former, like $\boldsymbol{S}_\Sigma$, has a finite correlation length which depends on $J_{out}/J_{in}$ ratio. The latter can be considered as disordered. Very short-range correlations between the normal components exists because the sum of these projections in all triangular clusters must be zero in the ground state, but the length of these correlations does not exceed the size of the clusters for any $J_{out}/J_{in}$ ratio. At the critical value of $J_{out}/J_{in} = 1.5$, where the transition to the three-dimensional ordered state occurs, the disordered component vanishes.

Another extreme case is when the in-plane interactions are negligibly small ($J_{in} \sim 0$). In this case, the long-range ordering along the $c$ axis is preserved, whereas correlations in the ($ab$) plane are completely lost. This situation can hardly be obtained in real systems and is therefore not of any practical interest.

## V. CONCLUSIONS

The lattice of $R$BaCo$_4$O$_7$ compounds, consisting of corner-sharing CoO$_4$ tetrahedra arranged in alternating Kagome and triangular layers, is an example of a new exchange topology exhibiting geometrical frustration. Superexchange paths with very strong antiferromagnetic interactions form bipyramidal clusters with high degree of frustration. These vertex-sharing clusters form chains running along the $c$ axis. In the ($ab$) plane, the chains are coupled via triangular units to create a three-dimensional network. Considering nearest neighbors only, there are four non-equal exchange paths in the system: two inside Kagome layers and two between Kagome and triangular layers. In compositions with hexagonal symmetry, all the exchange parameters are expected to be nearly equal. However, appropriate chemical substitutions and/or the presence of trigonal distortions, as found experimentally, can modify the ratio between the out-of-plane, $J_{out}$, and in-plane, $J_{in}$, interactions.

Within the mean-field approximation with an isotropic Heisenberg Hamiltonian, a long range ordering at finite temperature with propagation vector at the K-point of symmetry ($\boldsymbol{k}=\boldsymbol{a^*}/3 + \boldsymbol{b^*}/3$) takes place for $J_{out}/J_{in} > 0.7$. Below this value, the dominant Fourier modes are dispersionless and no selection of a particular wave vector is possible. The mean-field ordered state consists of collinear bipyramidal clusters arranged in the ($ab$) plane to make $120^o$ configurations. This spin structure becomes the ground state of the system also in the Monte-Carlo simulations for $J_{out}/J_{in} > 1.5$. Below this critical ratio, a quasi one-dimensional ordering of spins belonging to the triangular layers occurs along the $c$-axis. The correlations of these spins in the ($ab$) plane are short-range with correlation length dependent on $J_{out}/J_{in}$. For spins in the Kagome layers two components can be distinguished, one of them is correlated in the same way as the spins in the triangular sublattice, whereas another is disordered for any $J_{out}/J_{in} < 1.5$ The transition to the three-dimensional ordered state at $J_{out}/J_{in} = 1.5$ coincides with the disappearance of internal degrees of freedom in the bipyramidal clusters.

The obtained results clearly demonstrate that the behaviour of the system is very sensitive to the $J_{out}/J_{in}$ ratio. Varying the strength of trigonal distortions by doping or by the application of external pressure as well as by selective substitutions of cobalt ions in the triangular or Kagome sublattices, should be effective ways to vary this ratio and tune the nature of spin correlations in the system.

## Acknowledgment

The authors thank Prof. J. Rodriguez-Carvajal for providing the program ENERMAG. Work at Argonne supported under Contract No. DE-AC02-06CH11357 by UChicago Argonne, LLC, Operator of Argonne National Laboratory, a U.S. Department of Energy Office of Science Laboratory.